\newcommand{\be}{\begin{equation}}
\newcommand{\ee}{\end{equation}}
\newcommand{\ber}{\begin{eqnarray}}
\newcommand{\eer}{\end{eqnarray}}

\newcommand{\lsim}{\raisebox{-0.7ex}{$\stackrel{\textstyle <}{\sim}$ }}

\documentclass[preprint,epsf,aps,floats]{revtex4}
\usepackage{psfrag}
\usepackage{epsfig}
\begin{document}
\tighten
\begin{flushright}INT-PUB 04-19 \par UW/PT 04-10 \end{flushright}
\bigskip
\title{CFL Phase of High Density QCD at Non-Zero Strange Quark Mass}
\author{Andrei Kryjevski\footnote{abk4@u.washington.edu}}
\affiliation{Department of Physics and Institute for Nuclear Theory, University of Washington, Seattle, WA 98195}
\author {Daisuke Yamada\footnote{dyamada@u.washington.edu}}
\affiliation{Department of Physics,  University of Washington, Seattle, WA 98195}
\date{\today}

\begin{abstract}
We compute free energy of quark matter at 
asymptotically high baryon number density in the presence of non-zero strange
quark mass including dynamics of pseudo Nambu-Goldstone bosons due to chiral symmetry breaking, extending previously existing analysis
based on perturbative expansion in $m_s^2/4\mu\Delta.$
We demonstrate that the CFL$K^0$ state
has lower free energy than the symmetric CFL state for $0<m_s^2/4\mu\Delta<2/3$. We also calculate the 
spectrum of the fermionic quasiparticle excitations 
about the kaon condensed ground state in the regime $m_s^2/4\mu\Delta \sim 1$ and find that $(m_s^2/4\mu\Delta)_{crit}=2/3$ for 
the CFL-gCFL phase transition, the leading order result reported in \cite{Kryjevski:2004jw}, 
is not modified.

\end{abstract}

%\pacs{PACS numbers(s): ...}

\maketitle
\section{Introduction}

It is known that while Color-Flavor Locked (CFL) phase is the ground state of 3-color, 3-flavor massless QCD at 
asymptotically large baryon number density \cite{Alford:1998mk,Schafer:1999fe,Evans:1999at}, 
the ground state is modified by non-zero quark masses. An investigation of this issue has recently received a great deal of attention. 
See reviews \cite{Rajagopal:2000wf,Alford:2001dt,Nardulli:2002ma,Schafer:2003vz} and references therein. 

In the CFL phase with zero quark masses, quarks of all three 
flavors participate in the BCS pairing due to 
the attractive one gluon exchange potential in the ${\bar 3}$ channel.  The resulting condensate
spontaneously breaks the original $U(1)_B \times SU(3)_{Color} \times SU(3)_L \times SU(3)_R$ symmetry down to the diagonal subgroup, 
$SU(3)_{Color+L+R}$, causing the gauge bosons of the original $SU(3)_{Color}$ group to 
become massive by the Higgs mechanism \footnote{A local gauge symmetry
cannot really be spontaneously broken \cite{Elitzur:1975im}. We have to fix a gauge 
to define expectation values of gauge variant quantities.}. A linear combination of gluon 
and the photon remains 
massless; this unbroken gauge symmetry is referred to as 
$U(1)_{\tilde{Q}}$, a ``modified'' electromagnetism. 
The low energy bosonic excitations are parametrized by
$\Sigma = e^{2i\pi/f_{\pi}+i\,\eta^{'}/{\sqrt{6}\,f_{\eta^{'}}}}$ and $B=e^{i\,\beta/f_{B}}$ (with 
$\pi = \pi^a T^a$) where $\pi_a$'s are the pseudo-scalar octet
of Nambu-Goldstone bosons (NGB's) which arise from the breaking of chiral symmetry 
and $T_a$'s are the fundamental representation $SU(3)$ generators; $\eta^{'}$ and $\beta$ are NGB's 
due to breaking of the $U(1)_A$ (assumed to be a valid symmetry of 
the theory at high density) and $U(1)_{B}$ symmetries, respectively. 
The CFL symmetry breaking pattern is identical to that of QCD at low 
density \cite{Schafer:1998ef} with the important difference between the two cases being that at high density
the system is weakly coupled and the coefficients of the low energy EFT may be calculated from the underlying microscopic theory. 

In the presence of non-zero quark masses the ground state changes in the following way. A relatively large strange quark mass encourages 
the system to reduce the strange quark number density relative to the density of up- and
down-type quarks. The system responds to this stress by forming a condensate of collective excitations with quantum numbers of kaons in 
the ground state \cite{Schafer:2000ew,Bedaque:2001je,Kaplan:2001qk}. Kaon condensation allows the strange quark number density to 
be decreased without the costly breaking of Cooper pairs in the CFL background \footnote{In dense hadronic
matter, kaon condensation involves $K^-$ mesons, which introduce strange
quarks into the system \cite{Kaplan:1986yq} allowing one to approach $SU(3)$ neutrality
at high density. In contrast, the CFL$K0$ phase involves $K^0$ condensation
which reduces the net number of strange quarks participating in the BCS
pairing, driving the system away from $SU(3)$ neutrality in response to
a large strange quark mass.}. The $m_s^2/2\,\mu$ term, with $\mu$ being the quark number chemical potential,
acts as a dynamically generated chemical potential for strangeness; when it exceeds the kaon mass the Bose-Einstein condensation of $K^0$ 
is triggered and the CFL ground state undergoes a second order phase transition to the CFL$K^0$ phase where the parity and hypercharge are 
spontaneously broken by the condensate. Recently it was argued that taking into account ${\cal O}(\alpha_s\,m_s^2)$ terms in the effective 
lagrangian leads to more complicated phase diagram \cite{Kryjevski:2004cw}. However, in this work we neglect this effect for the sake of
simplicity.

So far the analysis has been performed to the leading order in $M^2/\mu \Delta,$ where $M={\mathrm diag}(m_u, m_d, m_s)$ is the quark mass 
matrix and $\Delta$ is the CFL gap parameter. It was shown that in the case of maximal kaon  
condensate (which occurs if we set $m_u =  m_d = 0$ and $m_s\neq 0$)
the free energy was lowered by $\delta \Omega = -f_{\pi}^2\,\left(m_s^2/2\,\mu\right)^2/2 \propto m_s^4$ while the hypercharge density 
in the CFL$K^0$ ground state was $n_Y=f_{\pi}^2\,\left(m_s^2/2\,\mu\right),$ where 
$f_{\pi}^2 = (21-8\,{\mathrm Log}\,2)\mu^2/36\,\pi^2$ is the leading order result for the pion decay constant
\cite{Son:1999cm,Bedaque:2001je,Kaplan:2001qk}. Extrapolating into the regime 
$m_s^2/4\,\mu\,\Delta \sim 1$ one may expect that the kaon condensation free energy is of the same order as the BCS 
condensation energy itself and the number of strange quarks involved in the condensate is of the same order as 
$3\,\mu^2 \Delta/{\pi}^2,$ 
the (approximate) number of strange quarks involved in the BCS pairing. Therefore, it may be important to take into account dynamics of 
NGB's when one
studies the properties of the ground state at large $m_s.$

Recently several calculations of dense quark matter free energy as a function of 
$m_s$ have been performed and CFL-gapless CFL (gCFL) phase transition was observed at $(m_s^2/4\,\mu\,\Delta)_{crit}=1/2$
\cite{Alford:2003fq,Ruster:2004eg,Alford:2004hz}. In the gCFL phase the quark-quark pairing still involves all colors and flavors, but the 
spectrum contains massless fermionic modes. Similar phases were discussed in the context of two flavor QCD \cite{Shovkovy:2003uu}.
The objective of this 
work is to obtain better understanding of the CFL ground state
as a function of $m_s$ which should help better understanding of the CFL-gCFL phase transition and of the properties of gCFL phase.

In Section 2 of this article we calculate free energy of the CFL$K^0$ phase to all orders in the
$m_s^2/4\,\mu\,\Delta$ expansion 
and show that the CFL$K^0$ persists 
as the energetically favored state even for $m_s^2/4\,\mu\,\Delta \sim 1.$ 
In Section 3 we extend the leading order calculation of the CFL phase quasiparticle excitation spectrum about the CFL$K^0$ ground state 
and observe that the result reported in \cite{Kryjevski:2004jw} is not modified by the higher order corrections. 

\section{Free Energy Calculation}

Let us first list some simplifying assumptions and approximations we make in this work
\begin{itemize}
\item{We neglect small light quark masses and assume $M={\mathrm diag}(0,0,m_s).$ We work in the regime 
${m_s^2/{\mu^2}} \ll 1,\,{m_s^2/{4\,\Delta \mu}} \leq 1$
so that the light-strange pairing is still possible \cite{Schafer:1999pb}. This means that we neglect deviation of 
$v_s = p_s/\mu = \sqrt{1 - m_s^2/\mu^2},$ the strange quark Fermi velocity, from 1. Note that in this regime the chiral expansion,
that is an expansion in ${\partial}/{\Delta}\sim{m_s^2/{\Delta \mu}},$ where one retains only few leading terms to 
attain desired accuracy, is expected to break down.
In this regime the relevant effective theory 
should include both bosonic excitations of diquark condensate and quasiparticles as the degrees of freedom. We will use mean field 
approximation in which we will freeze the ``radial'' oscillations of the condensate magnitude while allowing oscillations 
of the condensate around the CFL vacuum state corresponding to the NGB's due to the chiral symmetry breaking in the massless case. 
}
\item{Being in the weak coupling regime we will work to the leading order in $\alpha_s.$ Also we will neglect meson mass terms generated by 
quark masses. Quark mass terms connect particles and antiparticles and the resulting meson mass terms are generally suppressed by powers of 
$\Delta/\mu\propto {\mathrm Exp}(-{\mathrm const}/g_s)$ and/or $\alpha_s$ \cite{Son:1999cm,Schafer:2001za,Kryjevski:2004cw}.}
\item{We assume that $m_s$ is small enough so that we are in the CFL phase; we consider only color-flavor antisymmetric CFL gap term.}
\end{itemize}
For excitation energies below $\mu$ the relevant degrees of freedom are the nonet of quasiparticles 
and holes, ten pseudo NGB's due to the spontaneous symmetry breaking of global symmetries and soft gauge bosons. The corresponding 
effective theory is called the High Density Effective Theory (HDET) \cite{Hong:1998tn,Hong:1999ru,Bedaque:2001je,Schafer:2003jn}.
Let's consider the leading terms of the HDET Lagrangian of the CFL phase of high density QCD in the mean field approximation.
\ber
{\mathcal L} &=& 
- 2\times{{3 |\Delta|^2}\over{G}} + \nonumber \\ 
&+& {\mathrm Tr}\left [{L_v}^{\dag}( i v.\partial - (m_s^2/3\mu)
- {\mathrm \mu^L_s} - e\,v.A_{em}\,Q) {L_v}\right]
- g_s\,{\mathrm Tr}\left ({L_v}^{\dag}{L_v}v.A_c^{T}\right) + \nonumber \\ &+&
{{\Delta}\over{2}}\, \epsilon_{ijk}\epsilon_{abc}\, X^{k\,c}\, {L^{ia}_{-v\,\alpha}}{L^{jb}_{v\,\beta}}\,\epsilon^{\alpha\,\beta} + h.c.+
(L \leftrightarrow R,X \leftrightarrow Y,\mu_s^L \leftrightarrow \mu_s^R) -\nonumber \\ &-& 
\frac{1}{2}{\mathrm Tr}\,G^{\mu\nu} G_{\mu\nu}
- \frac{1}{4} F^{\mu\nu} F_{\mu\nu}.
\label{lagrangian1111}
\eer
Let us define all the quantities in (\ref{lagrangian1111})
\begin{itemize}
\item{The first term is the mean field potential with coupling constant $G$ being the coupling of the effective four fermion interaction 
generated by the hard gluon exchange.}
\item{
Fermionic fields
$L_v^{i\,a}(x)$ and $R_v^{i\,a}(x)$ are the left- and right handed quasiparticle fields that annihilate states with Fermi velocity 
$v^{\mu} = (1,\hat v), \,\hat v.\hat v = 1;$ flavor indices
$i,j,k$ take values $u,d,s;$ color indices $a,b,c$ take values $r,g,b;$ 
spinor indices $\alpha,\beta={\{}1,2{\}}$ are explicitly shown only in the gap term.
Under the original symmetry group $SU(3)_{c}\times SU(3)_{L}\times SU(3)_{R},$ 
$L_v$ transforms as $(3,3,1)$ and $R_v$ transforms as $(3,1,3).$ 
}
\item{
We separated the SU(3) singlet part of the leading HDET mass term $M^2 /2\mu.$ The singlet term will only 
be 
responsible for the baryon density shift due to non-zero quark masses, but will not affect the $m_s$ dependence of the condensation energy
and from now on will be neglected.
The remaining mass dependent terms are
\ber
\mu^L_s = \frac{M\,M^{\dag}}{2\mu} - \frac{1}{3}{\mathrm Tr} \frac{M\,M^{\dag}}{2\mu}, \nonumber \\
\mu^R_s = \frac{M^{\dag}\,M}{2\mu} - \frac{1}{3} {\mathrm Tr} \frac{M^{\dag}\,M}{2\mu}.
\label{mus}
\eer
With the $M$ we use $\mu^L_s =\mu^R_s =-\mu_s \,T_8,$ where $\mu_s = m_s^2/\sqrt{3}\mu.$ }
\item{
The NGB fields are collected in
$X^{kc}\sim\epsilon^{kij}\epsilon^{cab}\langle{L_{ai}^*L_{bj}^*}\rangle$ and 
$Y^{kc}\sim\epsilon^{kij}\epsilon^{cab}\langle{R_{ai}^*R_{bj}^*}\rangle,$ 
$3 \times 3$ unitary matrix valued composite 
fields describing oscillations of the left and right handed quark-quark condensates about the CFL ground state. 
Under the original symmetry group $SU(3)_{c}\times SU(3)_{L}\times SU(3)_{R},$ 
$X$ transforms as $(3,3,1)$ and $Y$ transforms as $(3,1,3)$ \cite{Casalbuoni:1999wu,Son:1999cm}. 
}
\item{$G^{\mu\nu}$ and $F^{\mu\nu}$ are the usual field strength tensors for the gauge fields $A^{\mu}_{c}$ and $A_{em}^{\mu}.$}
\end{itemize}

So, in our approximation the strange quark mass only appears in the role of an
effective chemical potential for strangeness. An 
infinitesimal $m_s$ induces kaon condensation in the ground state. What changes as $m_s$ increases? As was stated above, the kaon mass
terms generated by the quark masses are always suppressed by the powers of
$\Delta/\mu$ and/or $\alpha_s$ and cannot counterbalance the ``chemical potential'', ${m_s^2/{2 \mu}},$ driving the $K^0$ Bose-Einstein
condensation.
For as long as the CFL pairing pattern is preferred, there is nothing that can significantly alter the $K^0$ condensate in the ground 
state as $m_s$ increases. 
And in our approximation there is literally nothing that can change the form of the kaon condensate in the ground state. This is what our 
calculation will demonstrate.

Let us start by changing basis of quasiparticle fields to diagonalize the gap term. We write
\ber
{L_v} &=& X\,\psi_L = {1\over{\sqrt{2}}} \sum_{A=1}^{9} X \lambda^{A} {L^{A}_v},\nonumber \\
{R_v} &=& Y\,\psi_R = {1\over{\sqrt{2}}} \sum_{A=1}^{9} Y \lambda^{A} {R^{A}_v},
\label{q_chi}
\eer
where for $A=1,\ldots,8$ $\lambda^{ia}_{A}$ are the Gell Mann matrices 
and $\lambda^{i\,a}_{9} = \sqrt {2/3} \,\delta^{i\,a}$ \cite{Bedaque:2001je}.
The gap term becomes
\ber
{\mathcal L}_{gap} =
{1\over{2}} {\mathrm Det[X]} \Delta_{AB}{L^{A}_{-v\,\alpha}}{L^{B}_{v\,\beta}}\,\epsilon^{\alpha\,\beta} + h.c.
- (L\leftrightarrow R,{\mathrm X}\leftrightarrow{\mathrm Y})
\label{lagrangian_gap}
\eer
where $\Delta_{AB} = {1\over{2}} \Delta (\lambda^{aa}_{A} \lambda^{bb}_{B} - \lambda^{ai}_{A} \lambda^{ia}_{B}) = - \Delta, A,B = 1..8$ and
for $A,B = 9,$  $\Delta_{AB} = {1\over{2}} \Delta ({2\over{3}}\times 3 \times 3 - {2\over{3}}\times 3) = 2 \Delta$ and we used
$\epsilon_{abc} X^{am} X^{bn} X^{ck} = {\mathrm Det[X]}\epsilon^{mnk}.$ In this 
calculation we neglect $U_B(1)$ and $U_A(1)$ NGB's and set ${\mathrm Det[X]} = {\mathrm Det[Y]} = 1.$
The Lagrangian (\ref{lagrangian1111}) may be rewritten as 
\ber
{\mathcal L} &=& - 
2\times{{3 |\Delta|^2}\over{G}}+ \nonumber \\
&&
\sum_{A,B=1}^{9} {\Psi_{L}^A}^{\dag}{\left( \begin{array}{ccc}
%(p0 - p)
i\,v.\partial\,\delta^{AB} + {\mathcal X}^{A\,B}_{\hat v}
& \Delta_A \delta^{AB} \\
\Delta_A \delta^{AB} & 
%(p0 + p)
i\,\tilde{v}.\partial\,\delta^{AB} - {\mathcal X}^{B\,A}_{-\hat v} \\
 \end{array} \right)} {\Psi_{L}^B} + \nonumber \\
&&
\sum_{A,B=1}^{9} {\Psi_{R}^A}^{\dag}{\left( \begin{array}{ccc}
%(p0 - p)
i\,v.\partial\,\delta^{AB} + {\mathcal Y}^{A\,B}_{\hat v}
& -\Delta_A \delta^{AB}  \\
-\Delta_A \delta^{AB}  & 
%(p0 + p)
i\,\tilde{v}.\partial\,\delta^{AB} - {\mathcal Y}^{B\,A}_{-\hat v}
 \\
 \end{array} \right)} {\Psi_{R}^B} - \nonumber \\ &-& \frac{1}{2}{\mathrm Tr}\,G^{\mu\nu} G_{\mu\nu}
- \frac{1}{4} F^{\mu\nu} F_{\mu\nu},
\label{action}
\eer
where ${\tilde v} = (1,- \hat v)$ and $p = {\hat v}.{\vec p}-\mu.$ 
The rest of the definitions is as follows:
\be
\Psi_{L}^A =
\left( \begin{array}{ccc}
L^A_v \\
(L^A_{-v})^{\dag} \end{array} \right)
\label{psiL}
\ee
and
\be
\Psi_R^A =
\left( \begin{array}{ccc}
R^{A}_v \\
(R^A_{-v})^{\dag} \end{array} \right)
\label{psiR}
\ee
are the components of the Nambu-Gor'kov field;
\ber
{\mathcal X}^{A\,B}_{v} = {1\over{2}}{\mathrm Tr} \, \lambda_{A} X^{\dag} (i v.\partial  - \mu_s^L - e\,v.A_{em}\,Q) X \lambda_{B} -
{1\over{2}}\,g_s\,{\mathrm Tr} \, \lambda_{A} \lambda_{B} \,v.A_c^T,
\label{JXAB}
\eer
\ber
{\mathcal Y}^{A\,B}_{v} = {1\over{2}}{\mathrm Tr} \, \lambda_{A} Y^{\dag} (i v.\partial  - \mu_s^R - e\,v.A_{em}\,Q) Y \lambda_{B} -
{1\over{2}}\,g_s\,{\mathrm Tr} \, \lambda_{A} \lambda_{B} \,v.A_c^T.
\label{JYAB}
\eer

Before we proceed let us state the additional simplifying assumptions we make. 
\begin{itemize}
\item{We will neglect 
oscillations of the NGB fields about their ground state expectation values as the corresponding contribution to the free energy density is 
suppressed by $(\Delta/\mu)^2$ relative to the fermionic contribution. 
The values of $X$ and $Y$ in the ground state are to be 
determined by solving the corresponding equations of motion. 
}
\item{We will also neglect oscillations of the gauge fields about their ground state expectation values.
The $\langle A \rangle$'s are to be 
determined by solving the corresponding equations of motion which in the case of homogeneous quark matter are also the 
conditions for the gauge charge neutrality. These oscillations are important for the calculation of the gap 
parameter as the gap equation is dominated by the soft magnetic gluons \cite{Son:1998uk}, but here we do not attempt the calculation of 
the gap and
neglect the corresponding terms in the effective action as being of higher order in the coupling expansion. Our approximation is then 
equivalent to working in the model with a 
color-flavor antisymmetric short range interactions in the mean field approximation (NJL model) \cite{Alford:1998mk,Schafer:1999fe}. 
In our approximation the only role of the gauge fields is to ensure the 
gauge charge neutrality of the ground state \cite{Alford:2002kj,Kryjevski:2003cu}. 
The spatial components of gauge fields do not have sources and, therefore, vanish in the ground state. Then the kinetic terms for the 
gauge fields vanish, too.}
\end{itemize}

So, in this approximation we consider dynamics of quasiparticles in some gauge charge neutral meson field background.
To calculate free energy we need to determine what this background is.

Starting from (\ref{action}) let us formally integrate out fermions and then expand in powers of ${\mathcal X}$ 
and ${\mathcal Y}.$ This 
is the usual EFT expansion; effectively we expand in
$m_s^2/4 \mu \Delta \lsim 1$. Let us stress that we do not truncate this series at any finite 
order, that is, we do not rely on the smallness of the expansion parameter. We do rely on the analyticity of the free 
energy as a function of $m_s;$ our analysis is valid within the radius of convergence of the expansion and breaks down at the point of a 
phase transition.
We get the following effective action
%\pagebreak
\ber
{\cal S}[X,Y,A] &=& - \int_x 2{3 |\Delta|^2\over{G}} - {i} \,{\mathrm Tr}\, {\mathrm Log}\,{\mathrm S^{-1}} 
+\nonumber \\&+& \frac{i}{2} \sum_{n = 1}^{\infty} \,
\frac{1}{2\,n}
{\mathrm Tr}\left[ S\,{\left( \begin{array}{ccc}\tilde{\mathcal X}^{A\,B}_{v} & 0 \\
0 & -\tilde {\mathcal X}^{B\,A}_{-v}\\ \end{array} \right)}\,S\,{\left( \begin{array}{ccc}\tilde{\mathcal X}^{A\,B}_{v} & 0 \\
0 & -\tilde {\mathcal X}^{B\,A}_{-v}\\ \end{array} \right)}\right]^{2\,n} +\nonumber\\&+&
\frac{i}{2} \sum_{n = 1}^{\infty} \,
\frac{1}{2\,n}
{\mathrm Tr}\left[ S\,{\left( \begin{array}{ccc}\tilde{\mathcal Y}^{A\,B}_{v} & 0 \\
0 & -\tilde {\mathcal Y}^{B\,A}_{-v}\\ \end{array} \right)}\,S\,{\left( \begin{array}{ccc}\tilde{\mathcal Y}^{A\,B}_{v} & 0 \\
0 & -\tilde {\mathcal Y}^{B\,A}_{-v}\\ \end{array} \right)}\right]^{2\,n},
\label{expansion1}
\eer
where 
\ber
\tilde{\mathcal X}^{A\,B}_{v} = 
{1\over{2}}{\mathrm Tr} \, \lambda_{A}\left[ X^{\dag} (i v.\partial  - \mu_s^L - v.A_{em}\,Q) X - v.A_c^T \right]\lambda_{B} 
\label{JXABtilde}
\eer
and
\ber
\tilde{\mathcal Y}^{A\,B}_{v} = 
{1\over{2}}{\mathrm Tr} \, \lambda_{A}\left[ Y^{\dag} (i v.\partial  - \mu_s^R - v.A_{em}\,Q) Y - v.A_c^T\right] \lambda_{B}
\label{JYABtilde}
\eer 
and where ${\mathrm S^{-1}}$ is the inverse massless CFL propagator 
\ber
{\mathrm S^{-1}} = {\left( \begin{array}{ccc}
%(p0 - p)
i\,v.\partial\,\delta^{AB}
& \Delta_A \delta^{AB}\\
\Delta_A \delta^{AB} & 
%(p0 + p)
i\,\tilde{v}.\partial\,\delta^{AB} \\
 \end{array} \right)}.
\label{inverse_prop}
\eer
Also we have rescaled gauge fields to eliminate explicit dependence on the couplings. The operation ${\mathrm Tr}$ in (\ref{expansion1}) 
means trace over the 18 dimensional color flavor Nambu-Gor'kov space 
as well as over the space-time indices. 
The leading order terms in the expansion (\ref{expansion1}) produce leading order terms of the low energy EFT Lagrangian (for simplicity 
here we consider spatially uniform fields)
\cite{Zarembo:2000pj,Casalbuoni:1999wu,Son:1999cm}
\ber
{\cal L}&=& 
\frac{f_{\pi}^2}{2}\left[{\mathrm Tr} (X^{\dag}i D_0 X)^2 + {\mathrm Tr} (Y^{\dag}i D_0 Y)^2\right]+
{\cal O}\left(\mu^2\,\Delta^2\,(\frac{m_s^2}{\mu\,\Delta})^4\right),
\label{LXYAleadingorder}
\eer
where the covariant derivatives are given by
\ber
D_0 X &=& \partial_0 X + i X A^{c\,T}_{0} + i A_0^{em} Q X + i \mu_s^L X,\nonumber \\
D_0 Y &=& \partial_0 Y + i Y A^{c\,T}_{0} + i A_0^{em} Q Y + i \mu_s^R Y.
\label{covder}
\eer

For spatially homogeneous time independent configurations neutral under $U_{\tilde{Q}}(1),$
$\it{i.e.}$ for $[X,Q]=0$ and $[Y,Q]=0,$ it's easy to see that  
\ber
{\hat A} = A_0^{c\,T} + A^{em}_0\,Q = -\frac{1}{2}(X^{\dag}\mu_s^L X + Y^{\dag} \mu_s^R Y),
\label{Asolution}
\eer
the leading order solution, satisfies the equations of motion $\delta {\cal S}/\delta\,A_0 = 0$ for any term in the expansion 
(\ref{expansion1}) \cite{Kryjevski:2003cu}. 
Since the CFL$K^0$ is devoid of electrons, we set $A_0^{em}=0.$ If we constrain the ground state to be the symmetric CFL configuration and 
set $X=1,\,Y=1,$ then
\ber
{\hat A} = A_0^{c\,T} + A^{em}_0\,Q = - \mu_s \,T_8,
\label{AsolutionX=1}
\eer
and $A_{em}=0,\,A_c^3=0,\,A_c^8 = - \mu_s.$

The fact that the leading order solution ${\hat A}$ satisfies equations of motion at any order in the $m_s$ expansion follows from
the general form of gauge invariant effective action of a superconducting system
\cite{Weinberg:1986cq}. 

Let us integrate out color gauge fields which in this approximation is achieved by simply substituting (\ref{Asolution}) back into the 
effective action. Now we have
\ber
{\cal S}[\Sigma] &=& -2 \int_x {3 |\Delta|^2\over{G}} 
-\frac{i}{2}{\mathrm Tr}\,{\mathrm Log}\left[{\mathrm S^{-1}} + J_{V}+ J_{A}\right]\left[{\mathrm S^{-1}} + J_{V} - J_{A}\right],
\label{Schargenutral}
\eer
where
\ber
J_{V} ={1\over{4}}{\left( \begin{array}{ccc}{\mathrm Tr} \, \lambda_{A} [(X^{\dag} \mu_s^L X + Y^{\dag} \mu_s^R Y ), \lambda_{B}] & 0\\0& -
{\mathrm Tr} \, \lambda_{B} [(X^{\dag} \mu_s^L X + Y^{\dag} \mu_s^R Y ), \lambda_{A}]\end{array} \right)}
\label{JV}
\eer
\ber
J_{A} ={1\over{4}}{\left( \begin{array}{ccc}{\mathrm Tr} \, \lambda_{A} (X^{\dag} \mu_s^L X - Y^{\dag} \mu_s^L Y ) \lambda_{B} & 0\\0& -
{\mathrm Tr} \, \lambda_{B} (X^{\dag} \mu_s^L X - Y^{\dag} \mu_s^L Y ) \lambda_{A}\end{array} \right)}
\label{JR}
\eer

Considering only spatially homogeneous time independent configurations, we expand (\ref{Schargenutral}) in 
$m_s^2/4 \mu \Delta$ and obtain 
\ber
{S}[\Sigma] &=&  -2 \int_x{3 |\Delta|^2\over{G}} - {i}\,{\mathrm Tr}\, {\mathrm Log}\,{\mathrm S^{-1}} + 
i \sum_{n = 1}^{\infty} \,\frac{1}{2\,n}{\mathrm Tr}\left[{\mathrm S}\,J_{A}\,{\mathrm S}\,J_{A}\right]^{2\,n}.
\label{expansion2}
\eer
Here ${\mathrm Tr}$
also means trace over the 18 dimensional color flavor Nambu-Gor'kov space as well as over the space-time indices.
The leading term in the expansion corresponds to the leading term of the CFL $\chi$PT Lagrangian 
\ber 
{\cal L}_{{\chi}PT}=\frac{f_{\pi}^2}{4}\,{\mathrm Tr} \,D_0 \,\Sigma^{\dag} D_0 \,\Sigma +
{\cal O}\left(\mu^2\,\Delta^2\,(\frac{m_s^2}{\mu\,\Delta})^4\right),
\label{chipt}
\eer
where $\Sigma = X Y^{\dag}$ and 
$D_0 \,\Sigma = i \mu_s^L\,\Sigma - i \Sigma \,\mu_s^R,$ the Bedaque-Schaefer covariant derivative for 
a constant field \cite{Bedaque:2001je,Kaplan:2001qk}. 
We observe that $J_V$ terms are not present in the expansion.
One may show that 
\ber
\Sigma = \Sigma_{K^0}=\xi_{K^0}^2=
\left( \begin{array}{ccc}
1 & 0 & 0 \\
0 & 0  & i \\
0 & i & 0 \end{array} \right),
\label{sigmaK01}
\eer 
the leading order solution for the meson field \cite{Bedaque:2001je,Kaplan:2001qk}, satisfies 
equations of motion for $\Sigma$ for any term in
(\ref{expansion2}). 
\begin{figure}[t]
\centering{
\begin{psfrags}
\psfrag{y}{$\delta\,\Omega$}
\psfrag{x}{$m^2_s/4 \mu \Delta$}
\epsfig{figure=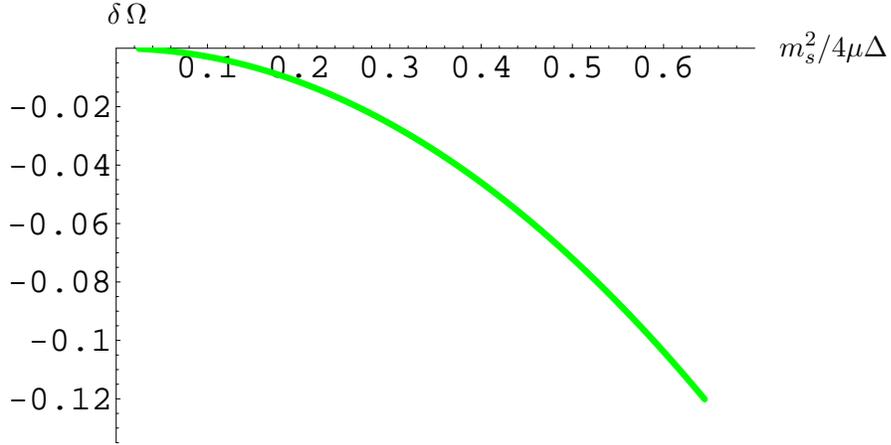, width=.65\textwidth}
\end{psfrags}
}
\caption{Free energy difference $\delta\,\Omega=(\Omega(\Sigma=\Sigma_{K^0}) - \Omega(\Sigma=1))/|\Omega^0_{CFL}|,$ between 
$\Sigma =\Sigma_{K^0}$ and $\Sigma = 1$ ground states as a function of $m^2_s/4 \mu \Delta.$ The massless CFL BCS condensation energy,
$\Omega^0_{CFL},$ is defined in (\ref{masslessCFL}). The 
$K^0$ condensed state has lower value of free energy.}
\label{domega}
\end{figure}
From $\Sigma$ we may determine values of matrices $X$ and $Y$ in the ground state which in the unitary gauge are 
given by $X_{K^0}=\xi_{K^0},$ $Y_{K^0}=\xi_{K^0}^{\dag}$ \cite{Casalbuoni:1999wu,Kryjevski:2003cu}.

\begin{figure}[t]
\centering{
\begin{psfrags}
\psfrag{y}{$n_{Y}/n_{BCS}$}
\psfrag{x}{$m^2_s/4 \mu \Delta$}
\epsfig{figure=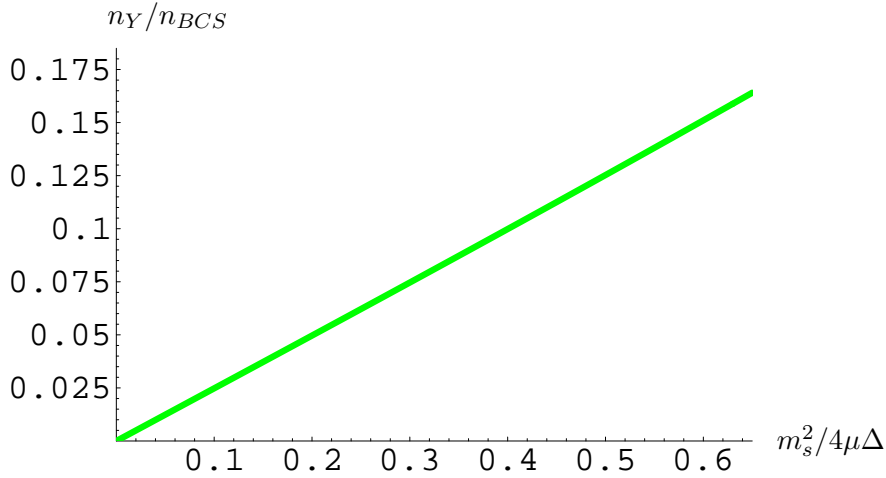, width=.65\textwidth}
\end{psfrags}
}
\caption{
The hypercharge density $n_Y/n_{BCS}$ in the CFL$K^0$ ground state
as a function of $m^2_s/4 \mu \Delta,$ where $n_{BCS}=3\,\mu^2\,\Delta/{\pi}^2,$ the approximate concentration of quarks of a given flavor
involved in the BCS pairing. The 
result is close to the leading order EFT prediction.}
\label{hypercharge}
\end{figure}

Now that we know the background gauge field values that ensure gauge charge neutrality and fields $X$ and $Y$ in the ground 
state, we may
calculate the difference in free energies between CFL$K^0$ and the $SU(3)$ symmetric CFL ground state.
The expression is
\ber
\delta\,{\Omega}&=& \Omega(A = {\hat A},\Sigma=\Sigma_{K^0}) - \Omega(A = {\hat A},\Sigma=1) = \nonumber \\
&=& \left[2\times{3 |\Delta|^2\over{G}} + 
2\,\pi \int_{p} \sum_{A=1}^{9}\,\left( \lambda^0_{A}(p;m_s) - \lambda_{A}(p;m_s,A = {\hat A},\Sigma=\Sigma_{K^0})\right)\right]
- \nonumber \\ &-&\left[2\times{3 |\Delta|^2\over{G}} + 
2\,\pi \int_{p} \sum_{A=1}^{9}\,\left( \lambda^0_{A}(p;m_s) - \lambda_{A}(p;m_s,A = {\hat A},\Sigma=1)\right)\right],
\label{Omega_difference}
\eer
where ${\{}p0-\lambda_{A}(p;m_s,A,\Sigma),\,p0+\lambda_{A}(p;m_s,A,\Sigma){\}}$ are the eigenvalues of the full inverse propagator in 
(\ref{action}) and  ${\{}p0-\lambda^0_{A}(p;m_s),\,p0+\lambda^0_{A}(p;m_s){\}}$ are the eigenvalues for the charge neutral normal phase. 
Note that we don't need an explicit expression for the latter. We neglect the difference in the 
value of the gap parameter for the two cases which we have checked to be small. Alternatively, we calculated the difference in free energy 
starting from (\ref{Schargenutral}). The results are
shown in Fig. {\ref{domega}}. The $K^0$ condensed configuration has lower free energy relative to the symmetric CFL state even for 
$m_s^2/4\mu\Delta \sim 1.$ The result of 
hypercharge density calculation in the CFL$K^0$ phase is shown in Fig. {\ref{hypercharge}}. We note that the numerical 
results are surprisingly close to the leading order EFT predictions.

In this calculation we performed the free energy renormalization by trading the bare coupling 
$1/G$ for the (physical) superconducting gap parameter $\Delta_0$ of the massless CFL by expressing $1/G$ from the gap equation 
\cite{Weinberg:1993dw}
\ber
{\delta {\Omega} \over{\delta \Delta}}\Biggl|_{m_s=0,\Delta = \Delta_0} = 0.
\label{gapeq_ms_zero}
\eer
Then, for example, the massless CFL BCS condensation free energy is given by \cite{Schafer:1999fe}
\ber
\Omega^0_{CFL}=-3\,\mu^2\,\Delta_0^2/{\pi}^2.
\label{masslessCFL}
\eer

\section{Quasiparticle spectrum}
Based on the results obtained, in this Section we extend the leading order calculation of the quasiparticle spectrum in the presence of 
kaon condensed background performed in \cite{Kryjevski:2004jw} into the 
regime $m_s^2/4 \mu \Delta \sim 1.$ Starting from (\ref{Schargenutral}) we write down the leading terms of the effective Lagrangian 
for the nonet of quasiparticle excitations about CFL$K^0$ ground state 
\ber
{\mathcal L} &=& {\mathrm Tr} \,{\psi}_{L\,v}^{\dag} iv.D \, {\psi}_{L\,v} + {\mathrm Tr}\, {\psi}_{L\,v}^{\dag}{\cal A}_0\, {\psi}_{L\,v} 
+ \frac{\Delta_{A\,B}}{2} {\psi}^{A}_{L\,-v}\,{\psi}^{B}_{L\,v} + h.c.+ \nonumber \\ 
&& {\mathrm Tr} \,{\psi}_{R\,v}^{\dag} iv.D \, {\psi}_{R\,v} - {\mathrm Tr}\, {\psi}_{R\,v}^{\dag}{\cal A}_0\,{\psi}_{R\,v}-
\frac{\Delta_{A\,B}}{2} {\psi}^{A}_{R\,-v}{\psi}^{B}_{R\,v} + h.c.,
\label{eff_lagrLR}
\eer
where the nonet field $\psi$ has been defined in (\ref{q_chi}).
This Lagrangian may be rewritten in the form prescribed by the chiral perturbation theory for the baryon fields \cite{Manohar:1996cq}
%\pagebreak
\ber
{\mathcal L} &=& {\mathrm Tr} \,{\psi}_{v}^{\dag} iv.D {\psi}_{v}+ 
\bar D\, {\mathrm Tr}\, {\psi}_{v}^{\dag} \gamma_5 \,{\{}v.{\cal A}, {\psi}_{v}{\}} + 
\bar F\, {\mathrm Tr} \,{\psi}_{v}^{\dag} \gamma_5 \,{[}v.{\cal A}, {\psi}_{v}{]} 
+ \nonumber \\ &+& \frac{\Delta_{A\,B}}{2} {\psi}^{A}_{-v}\,C\,\gamma_5\,{\psi}^{B}_{v} + h.c.,
\label{eff_lagr}
\eer
with $\bar D=\bar F=1/2$ and $C$ being the charge conjugation operator. 
The covariant derivative here is 
$D_{\mu}\,{\psi} = \partial_{\mu}\,{\psi} + i[{\cal V}_{\mu},{\psi}];$ the time components of the vector and axial currents modified by 
the quark mass terms are
\ber
{\cal V}_0=\frac{i}{2}\left(\xi^{\dag}\,(\partial_0 + i\,\mu_s^L)\,\xi + \xi \,(\partial_0 + i\,\mu_s^R) \,\xi^{\dag}\right) = 
-\frac{1}{2}\,\left(\xi^{\dag}\mu_s^L\xi + \xi \mu_s^R \xi^{\dag}\right),
\label{V}
\eer
and 
\ber
{\cal A}_0=\frac{i}{2}\,\left(\xi^{\dag}\,(\partial_0 + i\,\mu_s^L)\,\xi - \xi \,(\partial_0 + i\,\mu_s^R)\, \xi^{\dag}\right) = 
-\frac{1}{2}\,\left(\xi^{\dag}\mu_s^L\xi - \xi \mu_s^R \xi^{\dag}\right).
\label{A}
\eer

\begin{figure}[t]
\centering{
\begin{psfrags}
\psfrag{y}{$\min_p \lambda(p)/\Delta$}
\psfrag{x}{$m^2_s/4 \mu \Delta$}
\epsfig{figure=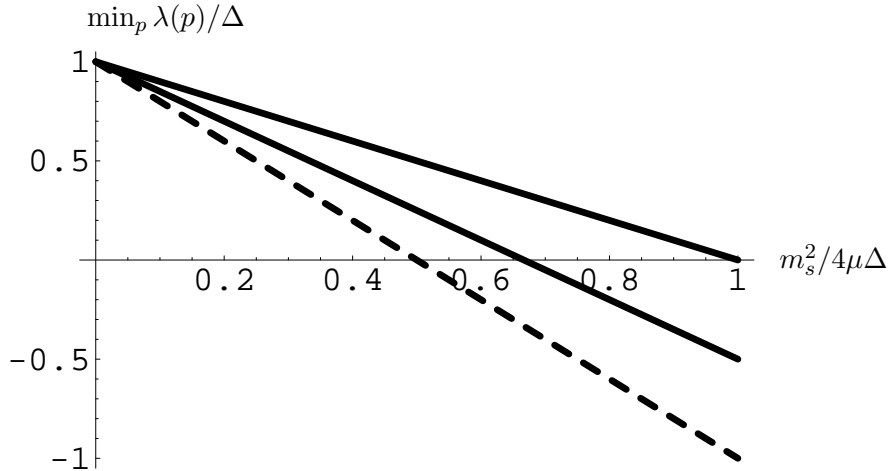, width=.65\textwidth}
\end{psfrags}
}
\caption{Masses of the quasiparticles that are lowered due to the presence of $m_s$ as a function of $m^2_s/4 \mu \Delta.$ The dashed line 
depicts a doubly degenerate $\Sigma=1$ case mode, solid lines are the
$\Sigma =\Sigma_{K^0}$ case modes. The results are in agreement with \cite{Kryjevski:2004jw}.}
\label{fig2}
\end{figure}

Shown in Fig. {\ref{fig2}} are the masses of the quasiparticles that become gapless as a function of $m_s^2/4 \mu \Delta$. 
The $\Sigma=1$ case develops gapless modes at $m_s^2/ 4 \mu \Delta = 1/2,$ while a CFL$K^0$ mode
becomes gapless at $m_s^2/ 4 \mu \Delta = 2/3.$ 
A detailed description 
%of this phenomenon 
is presented in \cite{Kryjevski:2004jw}. 
Here we only observe that the leading order in $m^2_s/4 \mu \Delta$ result and the result of the full calculation coincide. 
It follows from the fact that in the approximation we use the leading order in $m^2_s/4 \mu \Delta$ expressions for the background gauge fields and $\Sigma$ 
are not modified by the higher order terms. It has been argued that the appearance of a gapless mode signifies the point of CFL-gCFL phase 
transition \cite{Alford:2003fq}.
Our calculation confirms that due to the nontrivial flavor alignment in the ground state the CFL-gCFL phase transition is expected at 
$(m_s^2/ 4 \mu \Delta)_c=2/3$ which 
is 4/3 times higher than the symmetric CFL value. 

Note that a non analyticity at 
$m_s^{crit}$ is built into the expression for free energy (\ref{Omega_difference}) since at $m_s=m_s^{crit}$ the momentum integral over
$\lambda_A>0$ has to be modified to take into account the breach in the quasiparticle spectrum.
Then the value of 
$m_s^{crit}= \sqrt{8\,\mu \,\Delta/3}$ 
is the convergence radius of the $m_s$ series expansions (\ref{expansion1}), (\ref{expansion2}) which is consistent with the fact that 
at this point the system undergoes a phase transition.

We have neglected the quantum mechanical 
oscillations of the collective bosonic fields about CFL$K^0$ ground state.
One may have to worry about the stability of the ground state since as $m_s$ grows several quasiparticles are becoming lighter, whereas 
the massive mesonic excitations about CFL$K^0$ ground state have masses of order $m_s^2/2\,\mu$ and at 
some large enough $m_s$ a heavy Goldstone may decay into a particle-hole pair. However, 
due to the hypercharge and isospin charge conservation the quasiparticle-hole pair in the fermionic loop may not go on-shell for any $m_s$
and the ground state is stable.
\section{Discussion: some open questions}
We have demonstrated that taking into account dynamics of pseudo NGB's in the CFL phase of high density 
QCD and allowing a nontrivial flavor alignment in the ground state gives a sizable decrease in the free energy in the regime
$m^2_s/\mu \Delta \sim 1.$ It is remarkable that we were able to find a sensible approximation where it was possible to do the
calculation to all orders in the $m_s$ expansion. We also extended the leading order calculation of quasiparticle spectrum modified by the 
presence of kaon condensed background into the regime $m_s^2/4 \mu \Delta \sim 1$ and found the leading order result unmodified.

Several comments may be made based on the results of this work. First, since the CFL-gCFL phase transition is continuous, the 
kaon condensate in the 
gCFL ground state cannot be turned off abruptly and will persist at least for some values of $m_s>m_s^{crit}.$ While we argued that in the 
CFL nothing can significantly alter the magnitude of the kaon condensate, the non-zero non CFL gap terms, $\Delta_1$ and $\Delta_3$ 
defined in (\ref{gapnoncfl}),
may just do that by generating mass terms for the kaons. The question whether $K^0$ condensate will persist in the gCFL phase requires 
further investigation.

The most interesting issue that we have not addressed in this work is the investigation of the CFL-gCFL phase transition and of the gCFL 
ground state including dynamics of pseudo NGB's. 
In such a calculation one should allow deviations from the CFL pairing pattern in the gap term. For example, one may consider
\ber
\Delta^{m\,k} = \Delta_{CFL}\delta^{m\,k}+ \Delta_{13}^{m\,k},
\label{gapnoncfl}
\eer
with
\ber
\Delta^{m\,k}_{13} = {\mathrm diag} (\Delta_1,0,\Delta_3).
\label{gap13}
\eer
This pairing ansatz introduced by Alford $\it{et}$ $\it{al}$ in \cite{Alford:2003fq} is the simplest one that allows both disruption of the 
light-strange pairing 
($\Delta_3\neq0$) which is directly induced by $m_s,$ and of the pairing between light quarks ($\Delta_1\neq0$) which is due to the gauge 
charge neutrality and which, of course, is also indirectly induced by the strange quark mass. Unlike the CFL$K^0$ case, in the presence of 
the non 
CFL gap terms one 
does not know anymore the exact values of the background gauge fields and $\Sigma$ in the ground state.
A perturbation theory in 
$\Delta_{1,\,3}$ about the CFL$K^0$ state would seem to be a reasonable approach. To perform this calculation one needs to construct a low 
energy EFT in the regime $m_s^2/4 \mu \Delta \sim 1$
describing dynamics of light bosonic and fermionic modes.

In this simplified calculation we treated gluons as classical background fields. It would be interesting to repeat the calculation of the 
CFL free energy including the effects of quantum mechanical oscillations of the gluon fields into account.

These questions are left for future work. 
The full answer to the question: ``What is the ground state of dense QCD as a function 
of quark masses?'' is yet to come. 
\begin{center}
\large{\textbf{Acknowledgments}}
\end{center}
We wish to thank David Kaplan, Thomas Schaefer and Dam T. Son for helpful discussions along the way and commenting on the manuscript. The 
work 
is supported in part by the US Department of Energy grant DE-FG03-00ER41132 (A.K.) and by DOE contract No. DE-FG02-96ER40956
and by the Royalty Research Fund of the University of Washington (D.Y.).

\bibliography{cfl}% Produces the bibliography via BibTeX.
%  references
%

\end{document}